# Exoplanets and University-Industry Collaboration

Timothy Banks[1]  & Edwin Budding[2,3,4]

**Review**

A brief review is given of a university outreach programme by a commercial organisation, which uses the Kepler exoplanet data. Key insights derived from this research are presented, along with discussion of the benefits and challenges of such a collaboration between industry and academia. It is hoped that this account will be an inspiring example for others to emulate.

## Introduction

At first glance it might appear surprising that an international marketing research company, such as Nielsen, would be actively running an exoplanet research programme. Are they looking for new markets, perhaps?

Nielsen is a global measurement and data analytics company that provides measurement and insights into consumer behaviour and markets worldwide. For more than 90 years Nielsen has provided data and analytics based on scientific rigour and innovation, continually developing new ways to answer questions facing the media, advertising, retail and fast-moving consumer goods industries. It is this need to develop new insights, and methodologies to deliver those insights, that has led Nielsen to partner with research universities, developing and applying new methodologies. A particularly successful partnership has been developed over the last five years in Singapore, primarily with the National University of Singapore (NUS), but also with the Singapore Management University and Nanyang Technological University.

This paper gives a high level review of part of this larger university engagement programme by Nielsen. It concentrates on the use of publicly available datasets, which when combined with also easily available software, make for "bite sized" research projects by advanced undergraduate students. [5]

Joint research between universities and Nielsen in Singapore has covered a wide range of topics, including:
- Use of night time satellite imagery and measurement of 'light pollution' to measure urbanity in developing countries, leading to improvements in statistical sample designs developed by Nielsen (Tung et. al, 2014);
- Advanced statistical analysis to better understand verbal reporting of data (Olsen et al., 2017);
- Machine learning techniques applied to improve the efficiency of large scale sampling (in the published case, some 32,000 respondents), data collection and processing (Ming et al., 2015);

Nielsen is interested in sampling and statistical methods, being a company built on the collection, processing, and interpretation of data (as shown by the project list above), much of which are sourced from statistical samples. Optimisation is an area of particular interest, leading to improved allocation of resources, for example.

'Problems' industry can encounter with university engagement include whether data can be released outside the company and around intellectual property. For instance, there can be privacy concerns or worries about ownership of jointly developed intellectual property. It is often simpler to work with non-proprietary data sets and problems that are analogous to those being faced by a company such as Nielsen.

Use of the Kepler transit data addressed the two problems noted above, and provided an interesting optimisation problem. The methodologies and knowledge jointly explored (by the university and Nielsen) could be transferred by Nielsen internally to commercial problems, explaining the company's interest in the light curve datasets. Statistics and analytics are "hot" topics in Singapore, placing increasing demands on the universities as student numbers increase in this field. The universities are in strong need for co-supervisors and interesting research topics for the students, so the time was ripe for a fruitful collaboration between Nielsen and the Singapore universities. This paper outlines one such collaboration built over the last four years, with the National University of Singapore (in particular with the Department of Statistics and Applied Probability, see Figure 1) investigating exoplanets.

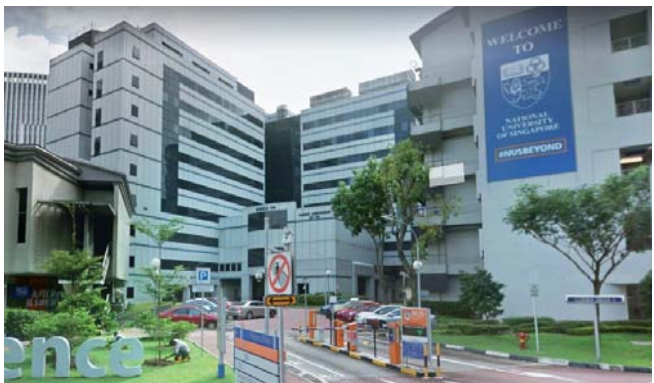

**Figure 1:** Google Streetview image looking towards the buildings housing the Department of Statistics and Applied Probability (National University of Singapore). Lecture halls are in the shorter building in the middle

[1] Data Science, Nielsen, 200 W Jackson Blvd #27, Chicago, IL 60606, USA
[2] School of Chemical and Physical Sciences, P. O. Box 600, Victoria University of Wellington, NZ
[3] Carter Observatory, 40 Salamanca Rd, Kelburn, Wellington 6012, New Zealand
[4] Department of Physics and Astronomy, University of Canterbury, Private Bag 4800, Christchurch 8020, New Zealand

[5] We note in passing that the same data sources have been used by the programme as the basis for a 'data science' competition. This is because Nielsen's interests in the outreach programme lie in two directions --- methodology development and talent acquisition. The latter has focused on increasing awareness of the Nielsen brand through activities such as sponsorship of data science and statistical programming competitions for large undergraduate groups (typically several hundred undergraduate students), case study competitions, research internships, and guest lectures at the universities.

## Methodology

The last two decades have seen an explosion in the number of planets confirmed to orbit stars other than the Sun. For further information on exoplanets, Pollacco et al. (2006), and Rice (2014) are excellent reviews. The transit method has been the leading technique for exoplanet detection, with the Kepler mission being the major contributor to the transit detections. Borucki et al. (2003) outline the scientific aims of this mission. The key feature of such an exoplanet transit is that the planet passes in front of its host star, blocking some of the star's light from reaching a given observer. Careful analysis of these light variations can provide information such as the exoplanet's radius and mass.

The Kepler Science Center manages the organization of data for scientific users. These data being easily accessible at the NASA Exoplanet Archive (*http://exoplanetarchive.ipac.caltech.edu/*, NEA), and available to researchers and the general public. The same website also provides useful online tools for the manipulation and preparation of the data for analysis.

Our research has centered on the uncertainties of parameter estimation for fits to the Kepler photometric data, in particular the transit regions of light curves. We believe it should be standard practice to state clearly uncertainties of light curve models and the associated "fits" to data, along with how uncertainties are estimated, as these give indications of the reliability of the estimates. This gains importance as the exoplanet field moves from the study of individual systems to population analysis. This interest in the determinacy of solutions explains why Nielsen partnered with a statistics department, in preference to a physics group.

The NUS students (see Figure 2) involved in the research are advanced undergraduate students, in their final year. The work makes up their Honours research thesis. While research novelty is not a requirement for the project, it is desired that the students demonstrate independent research skills and work at a sufficient standard that ideally their research can be published in the literature. NUS is one of the top universities globally (15th in the 2018 Quacquarelli Symonds and 22nd in the 2018 Times Higher Education world university rankings), requiring high standards from its students. The students have therefore been expected to build from first principles the fitting functions for the exoplanet transits, as well as the optimisation functions, before moving to comparison with other modelling codes (such as used in Budding et al., 2016; Eastman et al, 2013; and T. Barclay's fitting code [6]) and subsequently addressing research questions. The role of the authors has therefore been one of typical research supervisors, setting up the project problem and guiding the student through it.

The approach followed in the described research programme has been:

- Build from first principles a simple model for single-planet systems. As noted above, this is not novel, but an important step for the understanding of the students.

- Fit this model to simulated data sets, confirming that we could recover point estimates of the input parameters, and testing various optimisation techniques. This step gave confidence in these methods and their applicability to the research problem. Provided that the information content of the data is respected, the point estimates were effectively independent of the optimisation method, indicating a tidy convex optimisation problem

- Fit the model to `known' planetary systems, and compare the point estimates with those published in the literature. The goal of this step was to confirm that our model gave results in line with other researchers. At this step we also implemented Markov Chain Monte Carlo (MCMC) methods, to explore the determinancy of the fit (i.e., trying to understand the "error" around the point estimates for the model parameters). A Markov Chain is a random process where an object or process can move from one state to another with some underlying probability. We constructed a Markov Chain to find its stationary distribution, containing estimates of transit model parameters (see Figure 3 for example results of one system). Details on MCMC can be found in Brooks et al. (2011). We performed MCMC on data binned across many planetary orbits (reducing white noise and hopefully removing short term non-modelled effects), comparing the derived distributions with the previous results for each system. Finally, we validated our fits with those from the literature. The MCMC fits were in good agreement with the published results and also with the previous fits made in the project using other optimisation techniques. A key advantage of the MCMC modelling was the insight into how well 'determined' the point estimates of the model parameters are, and so we used only this methdology in the next step. As an aside, in future research we plan to investigate boot-strapping as an alternative method to explore parameter determinacy.

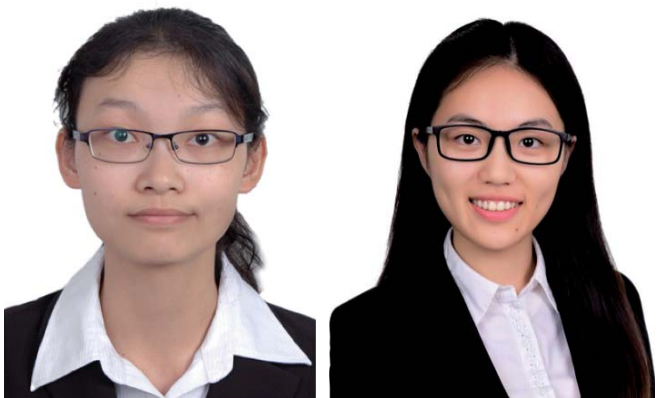

**Figure 2:** Two of the Honours students involved in the project. Ji Yi is on the left, and Huang Qing Ying on the right.

---

[6] Available for download at *https://github.com/mrtommyb/ktransit*

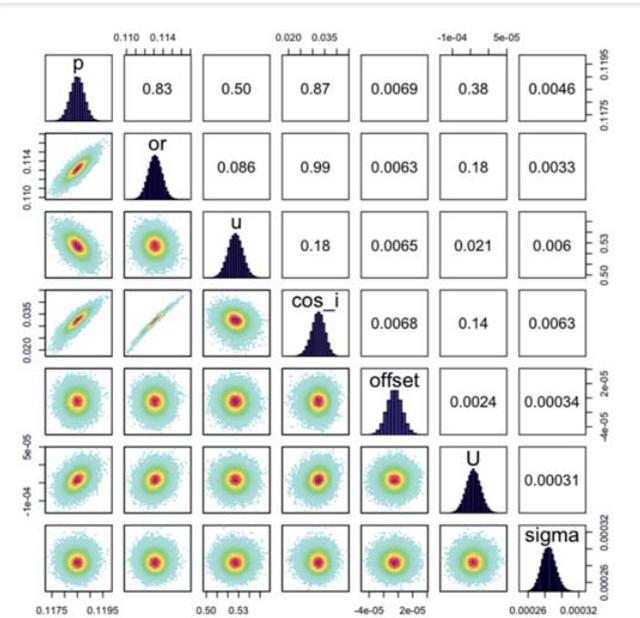

**Figure 3:** Pairwise correlation plot for Kepler 485, based on MCMC modelling using 4 chains, each of length 50,000 steps (post 'burn-in'). The density plots (in the lower left of the diagram) plot these 200,000 points for each parameter of the light curve model ("p" is the stellar radius, "or" the planetary radius, "u" the linear limb darkening, "cos_i" the cosine of the orbital inclination, "offset" the phase offset of the folded light curve, "U" the overall flux adjustment, and "sigma" the Gaussian noise in the binned data). The histograms along the diagonal show the "error" of the derived parameters (as well as the maximum likelihood), while the numbers in the upper right of the diagram are the correlation co-efficients between pairs of the parameters.

- Lastly, we applied the MCMC model to Kepler systems with no published results, partly so that the three Honours students involved in the project would be able to write and publish papers with new results. This has been successful: two research papers have been published (Budding et al., 2016; Ji et al., 2017) with two in preparation. Further details on the fitting model and results can be found in these first two publications. Sample high level results are given in the next section of this note.

The students have made use of the R (R Core Team, 2014) and Python programming languages, along with STAN (see Stan Development Team, 2016) for the MCMC implementation.

Other than the time the authors have spent in supervision, actual costs in this programme have been low. Substantial computing resources were not required for much of the work described in this note: the students and authors used laptops and open-source software. The tools and languages used (such as R, Python, and Julia) are of high popularity in data science today, making the projects even more appealing for the students as they see that they are learning "marketable" skills. For more intensive computing problems, access was available to internal Nielsen computing clusters and GPU (graphics processing unit) servers.

### Key results

Stellar and planetary radii, orbital periods, inclination (the orientation of the planet's orbital plane) and limb darkening co-efficients were derived for the previously unpublished systems Kepler 428, 485, 491, 706, 730, 760, 767, 802, and 824 in addition to confirmation for Kepler 1, 2, 5, 8, 12, 13, 20 and 77 (Mak, 2015; Yi, 2016; Huang, 2018).

A key question in this research was whether limb darkening could be derived from the data sets. We started with a simple linear model, frequently finding indeterminacy when even linear limb darkening was included as an optimisable parameter. The situation did not improve when a quadratic model was applied. We had wanted to test this model, as it is frequently used in the literature for exoplanet transit modelling. The subsequent MCMC optimisations confirmed that limb darkening is poorly defined from the data. The results also showed strong corellation with another parameter of key interest --- the radius of the exoplanet. We looked at the signal to noise ratio of the transit data, taking the depth of the transit as the signal, finding that the borderline S/N value for even a linear limb darkening coefficient to be determinable is about 15. For noisy data with low S/N, it is preferable to take the limb darkening coefficient calculated from appropriate stellar atmosphere models rather than derive it from the data. This work confirmed the earlier and independent study of Csizmadia et al. (2013) into the determinability of limb darkening coefficients. It also points to the importance of 'error analysis' in such light curve modelling as described in this paper. A surprising number of exoplanet transit papers do not report 'uncertainties' in their derived parameters from transit model fits, while others appear to be fitting complicated models beyond the information content of the data (overfitting the data).

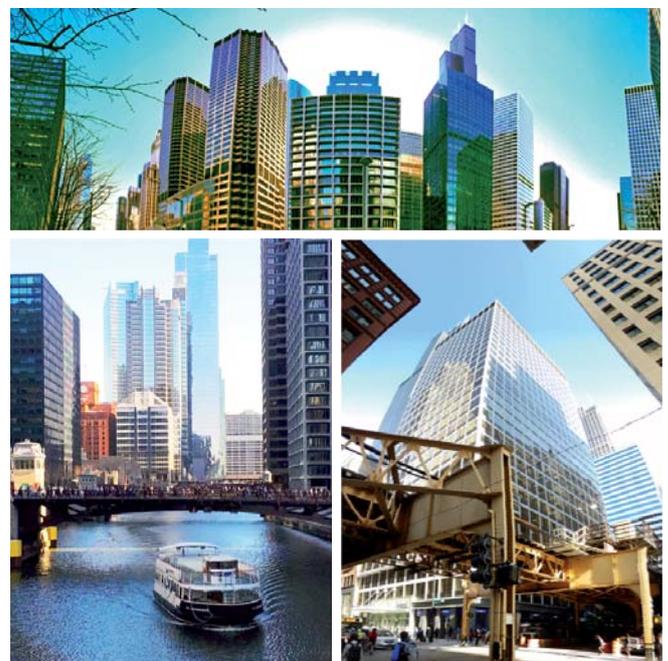

**Figure 4:** Views of Chicago. The top photograph shows buildings along the Chicago River (part of the "The Loop"). The Nielsen office is behind the tall black building on the right. The lower right photograph is a photograph of the Nielsen Chicago office, with an elevated railway in front of the building. Nielsen occupies roughly half of this building. The photograph on the lower left is looking up the Chicago River, some two minutes walk from the Nielsen office (photographs by the authors).

A second key question centred around the impact on derived parameters of the integration period of the data set. Kepler data are available in two cadences, short cadence and long cadence. Each cadence is composed of multiple 6.02-s exposures with associated 0.52-s readout times (Gilliland et al., 2010). There is a longer time interval between observations for long cadence data (30 minutes) as compared to their short cadence (1 minute) counterparts. Later Kepler data sets are often only available on the NEA in long cadence. Kepler 491 had both long and short cadence data sets available, optimisations showed statistically significant differences in the derived planetary radii and system inclinations depending which integration periods were used. The long cadence data sets led to larger planet radii estimates and systems tipped further from the line of sight. This was confirmed through analysis first of Kepler 12, and then through binning tests of short cadence data. This point is of concern given the lack of short cadence data for many systems, and the increasing interest in population studies. Even with formal MCMC modeling uncertainties for parameters, the implicit binning in long cadence data can lead to incorrect estimates for the parameters and their "errors". The impact will vary across systems, primarily driven by the ratio of the transit time to the integration time. We believe this is a serious concern for such population work. This work will be published in Huang et al. (2018), where further details might be found.

Knowledge of such parameter distributions will be important for any `meta-analyses' based on collection of many individual and separate studies deriving exoplanet parameters. Without a clear understanding of confidence limits in modeling results and also the possible biases introduced by the implicit binning in datasets by the exposure times, meta-analyses might reach incorrect conclusions by placing undue weight on `observations' of dubious confidence or indeed the reverse. We strongly encourage other researchers in the field to use techniques such as MCMC and boot-strapping to explore the determinacy of their model fit and the confidence that should be placed in the parameter estimates.

## Discussion

Publically available astronomical data sets can allow students access to high quality observations, such as with Kepler. This can allow the students to make valuable contributions to the scientific literature, make a start along careers in science, and feel engaged with the cutting edge of the field. The work by the NUS students has been of sufficient scientific interest that it has been successfully published in research journals. Students can feel connected with the frequent news items in the media on exoplanets, and "feel the excitement" of working as active researchers in a developing field.

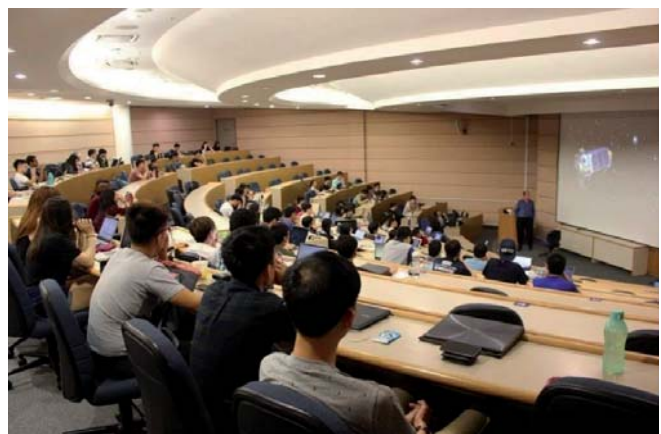

**Figure 5:** Guest lecture on exoplanets and model building by one of the authors at the Department of Statistics and Applied Probability's main lecture theatre

Astronomy has a wide appeal, and is an excellent gateway subject to build interest in science and technology --- indeed, the authors have used Kepler data for a data analytics competition they arranged and sponsored at the National University of Singapore, with over one hundred students competing during a weekend in a "hackathon" to derive parameters for a transiting exoplanet system. Such a strong turnout speaks to the appeal exoplanets have for the general public and students (see also Figure 5 for a picture taken of a guest lecture on exoplanets at NUS by one of the authors).

As noted above, such partnerships can be a "win-win" scenario for industry and universities alike --- providing interesting projects and supervision for research students, and knowledge transfer and development for companies. The students also add capacity for the companies, allowing interesting research problems to be explored which otherwise might not have been, given competing priorities and "limited bandwidth" for permanent employees.

All in all, the authors recommend such collaborations, particularly as they are a way of building interest in astronomy and science. While it takes a commitment from the supervisors, as you will be responsible to provide support and development opportunities for the students, we strongly recommend such engagement with universities when such opportunities are available. We note that we see high value from our current partnership. Indeed, despite one of the authors (TB) being relocated outside Singapore for a few years (to Chicago, see Figure 4), we have carried forward this productive relationship described above using contemporary internet communication facilities.

## Final Comments

This note has outlined an active research collaboration between a university and a commercial organisation, in a field that at first glance might seem a surprising topic for such collaboration. It has discussed the benefits to both organisations in the collaboration, as well as to the students who are able to work on 'live research'. Some example results have been discussed, which have been sufficiently novel to be published in the research literature.

We have tried to demonstrate that publicly available datasets, when combined with supervision and easily available software, can make for "bite sized" research projects for students or even as the basis for 'data science' analytics competitions.

We hope that this note might encourage some readers to engage with their local education organisations to increase interest in astronomy, science, and research. While the programme work described here has centred on that of advanced undergraduates, perhaps there are opportunities for scaled engagement with advanced secondary school students on smaller projects? As an example, the American Association of Variable Star Observers (AAVSO) has reached out to high school and tertiary students, together with their instructors, who wish to conduct and publish astronomical research. The group offers AAVSO data, suggestions for projects, and software as well as a possible journal for publication (see Percy, 2017). Data analytics competitions similar to those mentioned above could be another avenue. The authors offer no direct suggestions or recommendations on how such a programme could (or even should) be extended to the general public; based on our experience formerly working in public outreach at Carter Observatory we see this as something with its own challenges and outside the scope of this paper.

By engaging with students via carefully selected and managed data and analytic challenges, we believe that it could lead to increased interest in astronomy (and science in general). We also believe by taking such active steps into the community, astronomers can "prepare the fields for the future".

## Acknowledgements

This research has made use of the NASA Exoplanet Archive, which is operated by the California Institute of Technology, under contract with the National Astronautics and Space Administration under the Exoplanet Exploration Program. We are particularly grateful to Professor Lim Tiong Wee (Deputy Chair, Department of Statistics and Applied Probability, NUS) for his strong and ongoing support of this research programme, as well as to Professor Emeritus Michael D. Rhodes (Bingham Young University). The authors also thank the Economic Development Board of Singapore for partial support for this project, and Dr. N. Rattenbury for his helpful comments as referee for this paper.

## References

- Brooks, S., Gelman, A., Jones, G., & Meng, X., (2011), Handbook of Markov Chain Monte Carlo, Chapman and Hall, CRC.
- Budding, E., Rhodes, M. D., Püsküllü, C., Ji, Y., Erdem, A., & Banks, T., (2016), *Astrophysics and Space Science*, **361**(10), p. 346.
- Csizmadia, Sz., Pasternacki, Th., Dreyer, C., Cabrera, J., Erikson, A., & Rauer, H., (2013), *A&A*, **549**, A9.
- Eastman, J, Gaudi, B., & Agol, E, (2013), *PASP*, **125**, 83.
- Gilliland, R. L., Brown, T. M., Christensen-Dalsgaard, J., et al., (2010), *PASP*, **122**, 131.
- Huang, Q. Y., (2018), Unpublished Honours Thesis, Department of Statistics and Applied Probability, National University of Singapore.
- Huang, Q. Y., Budding, E., Puskullu, C., Rhodes, M.D., & Banks, T., (2018), *Astrophysics and Space Science*, in preparation.
- Ji., Y., (2016), Unpublished Honours Thesis, Department of Statistics and Applied Probability, National University of Singapore.
- Ji, Y., Banks, T., Budding, E., & Rhodes, M.D., (2017), Astrophysics and Space Science, 362(6), p. 112.
- Mak, F.H., (2015), Unpublished Honours Thesis, Department of Statistics and Applied Probability, National University of Singapore.
- Ming, S.S., Tung, W. L., & Banks, T., (2015), in "Natural Computation (ICNC), 2015 11th International Conference on", IEEE, doi: 10.1109/ICNC.2015.7378001, p. 261.
- Olsen, K., Li, X. Y., & Banks, T., (2017), *International Journal of Market Research*, **59**(3), p. 301.
- Percy, J.R., 2017, IAU Division C1 (Astronomy Education and Development), Newsletter 86, 19.
- Pollaco D.L., & 27 authors, (2006), *PASP*, 118, p. 1407.
- R Core Team, (2014), "R: A language and enviroment for statistical computing", R Foundation for Statistical Computing, Vienna, Austria.
- Rice, K., (2014), Challenges, 5, p. 296.
- Stan Development Team, (2016), "Rstan: the R interface to Stan", *http://mc-stan.org/* .
- Tung, W. L., Zhao, J. Y., & Banks, T., (2014), in "Asia-Pacific 2014 – Celebrating Asian Creativity", p. 1, ESOMAR Publication Series Vol S364 APAC, ISBN: 92-831-0273-8.

## Appointed Officers

| | | | |
|---|---|---|---|
| Electronic Newsletter Editor: | Mr A C Gilmore | Southern Stars Editor: | Mr R W Evans |
| Web Master: | Mr P Jaquiery | Archivist: | Mr G Hudson |
| Property Officer: | Mr G Hudson | Membership Secretary: | Ms M Head |
| Hon. Solicitor: | Mr J McCay | Hon. Auditor: | Mr A Wheelans |

## Sections

| | | |
|---|---|---|
| Astrophotography | Director | Mr J Green, 374 Sunnyside Road, RD2, Albany 0792, NZ. |
| | | *http://www.rasnzaps.co.nz, jonathangreen@xtra.co.nz* |
| Comet and Meteor | Director | Mr J Drummond, P O Box 113, Patutahi 4045, NZ. |
| | | *http://www.cometeor.co.nz* |
| Dark Skies Group | Convenor | Mr S C Butler, 30 Hoffman Court, Invercargill 9810, NZ. |
| | | *http://www.rasnz.org.nz/groups-and-sections/dark-skies-group* |
| Occultation | Director | Mr S R Kerr, 22 Green Ave, Glenlee, Queensland 4711, Australia. |
| | | *http://www.occultations.org.nz* |
| Professional Astronomers' Group | | Dr N. Rattenbury, Department of Physics, University of Auckland, NZ |
| | | *http://www.rasnz.org.nz/groups-and-sections/professional-astronomers-group* |
| Space Weather | Director | Damian McNamara, 16 Harlech St, Oamaru 9400, NZ. |
| | | *solaur.science@gmail.com* |
| Variable Stars South | Director | Mr M G Blackford, 25 Bambridge St, Chester Hill, NSW 2162, Australia. |
| | | *http://www.variablestarssouth.org* |

## Fellows

| | | | | |
|---|---|---|---|---|
| Mr W H Allen | Prof E Budding | Mr S C Butler | Dr G W Christie | Mr R W Evans |
| Mr A C Gilmore | Prof. J B Hearnshaw | Ms P M Kilmartin | Mr B R Loader | Ms J M McCormick |
| Ass. Prof. K R Pollard | Dr D J Sullivan | Mr W S G Walker | Prof. P C M Yock | |

## Honorary Members
Gerry Gilmore, FInstP, ScD, MAE, FRS
Thomas Richards MA(Hons VUW), DPhil(Oxon)
Brian Warner BSc(Hons), PhD, DSc(London), MA, DSc(Oxon), Assoc RAS, FRSSA$_f$, MASSA$_f$